\documentclass[12pt,onecolumn,showpacs,preprintnumbers,amsmath,amssymb,amsthm,mathrsfs]{revtex4}
\usepackage{color}
\usepackage{graphicx}% Include figure files
\usepackage{dcolumn}% Align table columns on decimal point
\usepackage{bm}% bold math

\begin{document}

\title{Liquid behavior of hot QGP in the finite temperature field theory}

\author{Hui Liu}
 \email{tliuhui@jnu.edu.cn}
\affiliation{%
Physics Department, Jinan University, Guangzhou(510632), P.R.China
}%
\author{Defu Hou}

\author{Jiarong Li}

\affiliation{%
Institute of Particle Physics, Central China Normal University,
Wuhan(430079), P.R.China
}%
\date{\today}% It is always \today, today,

\begin{abstract}
In this paper, we compare the dispersion relations of hard thermal
loop and complete one loop. It is shown that in the dynamical
screening regime, the completely one-loop calculation presents a
prominent threshold frequency, below which no pure imaginary mode
survives. This phenomenon is responsible for the oscillatory
static in-medium potential and ultimately results in a damping
oscillation of the radial distribution function. We consider this
typical shape is the footprint of liquid QGP.
\end{abstract}


\maketitle

\section{Introduction}

The experiments of ultra-relativistic heavy ion collision at RHIC
provide us a platform to study the quark-gluon plasma(QGP) signal
as well as its novel properties. One of those surprises the
scientists is the low viscous flow. At Au+Au 200GeV collision, the
elliptic flow $v_2$ can be well fitted by an ideal hydrodynamics
up to 2GeV of the transverse momenta\cite{star,phenix}, which
implies a perfect fluid behavior. This perfect behavior of QGP
makes people consider it in a liquid state\cite{Thoma2,Peshier},
with the temperature slightly above the critical temperature
$T_c$. How to understand such a good liquid of QGP is a
fundamental problem that attracts much attention. Some ideals and
methods came from other fields, for example the AdS/CFT
correspondence from the superstring theory and the physics of
strongly coupled QED plasma. For more details, please refer to the
report of E. Shuryak in Ref\cite{Shuryak} and references therein.

In this paper, we try to investigate the radial distribution
function of liquid QGP in the framework of finite temperature
theory. Hard Thermal Loop(HTL) approximation and HTL resummation
scheme were widely used in thermal field theory when discussing
measurable medium effects such as Debye screening, collective
modes, particle energy loss so on and so forth. The HTL physics
were proved reliable in the temperature limit. For example, it can
represent the correct collective modes in hot
plasma\cite{LeBellac}. The boson and fermion damping rates
obtained in the HTL resummation scheme are positive and gauge
invariant even in the non-abelian system\cite{Pisarski,Braaten}.
However, although the HTL has this and that good qualities, it has
its own restrictions. The HTL approximation as well as the
corresponding resummation scheme request the high temperature
limit which is not a trivial condition for a real system like the
QGP at $1\sim 2T_c$. This temperature is obvious not reaching the
high temperature limit so that the HTL scheme might be doubtful.
To avoid such suspicion, one can adopt complete one loop scheme
instead of HTL.

In this paper, we will start with QED plasma, comparing the
dispersion relations of HTL and complete one loop, demonstrating
their distinct screening behaviors. Then we will turn to the
quark-gluon plasma, calculating the static in-medium inter-quark
potential and the radial distribution function. The damping
oscillatory radial distribution function suggests the QGP might be
in a liquid state. Finally, we will discuss the general factors
that decide the state of matter, pointing out a possible way to
study the properties of QGP liquid.

\section{dispersion relation}

Dispersion relation is a basic relation of many-particle system
which carries essential physical information. A slight difference
between dispersion relations may indicate totally different
physics. In this section, we will compute the QED dispersion
relations at HTL and completely one-loop level respectively. One
will see the distinct dispersion curves in both dispersion regime
and dynamic screening regime.

The dispersion relation is defined as the energy-moment relation
at the pole of full boson propagator,
\begin{eqnarray}\label{define dispersion}
&&\omega^2-q^2-\Pi_{L}(\omega,q)=0 \\
&&\omega^2-q^2-\Pi_{T}(\omega,q)=0
\end{eqnarray}
where $\Pi_{L}(\omega,q)$ and $\Pi_{T}(\omega,q)$ are the
longitudinal and transverse components of boson polarization
tensor respectively. In this paper we just take the longitudinal
dispersion relation as an example and study the color-electric
properties of hot plasma.

In HTL approximation,
\begin{equation}\label{HTL self-energy}
\Pi_{L}^{\mbox{\tiny HTL}}(\omega,q)=-\frac{4\pi \alpha
T^2}{3}\left[1-\frac{\omega}{2q}\ln\left(\frac{\omega +q}{\omega
-q}\right)\right]
\end{equation}
where $\alpha=1/137$ is the fine structure constant of QED. For a
complete one loop,
\begin{eqnarray}\label{one loop self-energy}
\Pi_{L}^{\mbox{\tiny
one-loop}}(\omega,q)&=&\frac{4\alpha}{\pi}\int^\infty_0 dp
\frac{p^2
n_f}{E_p}\left[\frac{\omega^2-q^2+4E_p^2+4\omega E_p}{4pq}\ln\left(\frac{\omega^2-q^2+2\omega E_p+2pq+i\epsilon}{\omega^2-q^2+2\omega E_p-2pq+i\epsilon}\right)\right.\nonumber\\
&&+\left.\frac{\omega^2-q^2+4 E_p^2-4\omega
E_p}{4pq}\ln\left(\frac{\omega^2- q^2-2\omega
E_p+2pq-i\epsilon}{\omega^2-q^2-2\omega
E_p-2pq-i\epsilon}\right)-2\right],
\end{eqnarray}
where $E_p=\sqrt{p^2+M^2}$, and $M$ is the electron mass.
$n_f(E_p)=(e^{\beta E_p}+1)^{-1}$ is the Fermi-Dirac distribution
function with $\beta=1/T$.

\begin{figure}
   \resizebox{10cm}{!}{\includegraphics{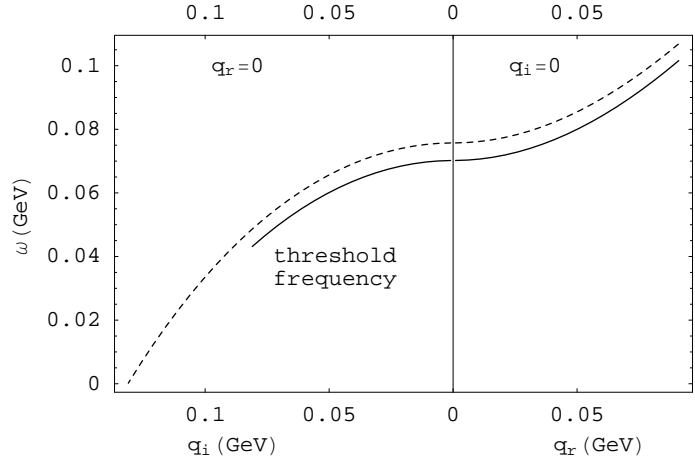}}
   \caption{Comparison of dispersion relations between HTL and completely one-loop calculations. The dashing line denotes for the HTL
            calculation and the solid line is for the completely one-loop calculation.\label{fig disp_comp} }
\end{figure}

Inserting Eqs.(\ref{HTL self-energy}) and (\ref{one loop
self-energy}) into Eq.(\ref{define dispersion}) and figuring out
the relation between $\omega$ and $q$ numerically, one could
obtain FIG.\ref{fig disp_comp}. This figure is plotted in not only
the dispersion regime where the momentum $q$ is real, but also the
dynamic screening regime where $q$ is pure imaginary. In
FIG.\ref{fig disp_comp} the abscissa combines both regimes,
separated by a zero line of $q=0$. The right area to the zero line
is for the common dispersion relation when the momenta are real.
The left area, on the contrary, is the dynamic screening regime
for pure imaginary momenta.

The HTL dispersion relation has been obtained and discussed in
details\cite{Weldon}. We represent it in FIG.\ref{fig disp_comp}
with dashed curves to compare with the complete one loop. However
we do not intend to compare the whole regime, since the two curves
in the normal dispersion regime behaves very similar. Instead, we
would like to concern about the prominent difference in the
dynamical regime. In this regime, the HTL curve reaches the
abscissa, indicating a screening effect at zero frequency referred
to the well-known Debye screening. While in the completely
one-loop case, an threshold frequency shows up, below which no
pure imaginary mode survives. That is to say a real part of the
momentum is necessary and the dynamical screening described by the
HTL \cite{Weldon} is broken up. Especially, in the static limit
where $\omega\rightarrow 0$, the plasma is not screened with Debye
form contributed by the pure imaginary mode. Instead, the
screening oscillates due to the complex mode in the completely
one-loop calculation. We will see it in the next section.

\section{Oscillatory potential}

So far the Debye screening picture has been changed in the
completely one-loop calculation based on the dispersion analysis
in last section, one would like to check the static potential and
see how it will look like in the new picture.

\begin{figure}
%\begin{minipage}[t]{0.5\linewidth}
 %\centering
 \begin{center}
   \resizebox{7cm}{!}{\includegraphics{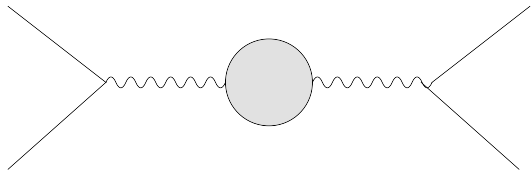}}
 \end{center}
\caption{\label{fig potential_diag}Diagrammatical description of
in-medium interparticle potential. The shadowed circle stands for
all possible polarization patterns.}
%\end{minipage}%
\end{figure}

In the relativistic plasma, the in-medium  potential is explained
by the skeleton diagram with full boson propagator, as shown in
FIG.\ref{fig potential_diag}. The shadowed circle denotes all
possible polarizations. In math language, it is
\begin{equation}\label{potential}
V(r)=\frac{\alpha}{\pi r}\
\mbox{Im}\int^\infty_{-\infty}dq\frac{qe^{iqr}}{q^2-\Pi_L(0,q)},
\end{equation}
where $r$ is the distance between two arbitrary electrons. To
perform the integral in Eq.(\ref{potential}), one should construct
a contour according to the analytic structure of the integrand,
locating all poles within the contour on complex plane. We would
like to point out here that the Eq.(\ref{potential}) is actually
involving a resummation scheme, because the effective boson
propagator is obtained from Dyson-Schwinger equation.

To demonstrate the general form of the potential contributed by
poles, one can first define the pole as
\begin{equation}
q_{\mbox{\tiny pole}}=q_r+iq_i,
\end{equation}
where $q_r$ and $q_i$ are the real and imaginary parts of the
pole. With this definition, one can perform the contour integral
and find
\begin{equation}\label{osc potential}
V(r)=\sum_{\mbox{\tiny poles}}\frac{2\alpha}{a^2+b^2}\frac{e^{-q_i
r}}{r}\ [a\cos (q_r r) +b\sin ( q_r r)],
\end{equation}
where the sum includes all pole contributions.  $a$ and $b$ are
defined as the real and imaginary parts of the residue,
\begin{equation}\label{ab}
\left.\frac{(q^2-\Pi_L)'}{q} \right|_{q=q_r+i q_i}=a+i b,
\end{equation}
with the prime denoting $\partial/\partial q$.

Notice that the pole of the integral in Eq.(\ref{potential}) is
nothing else but the point of $\omega=0$ on the dispersion curve.
Due to the appearance of the threshold frequency in FIG.\ref{fig
disp_comp}, the potential from HTL polarization and completely
one-loop polarization may behave differently. The HTL dispersion
curve extends directly to zero frequency in the dynamic screening
regime, which means the pole is purely imaginary with $q_r=0$ at
the static limit. More explicitly,
\begin{equation}
\Pi_L^{\mbox{\tiny HTL}}(\omega\rightarrow 0,q)=-\frac{4\pi\alpha
T^2}{3},
\end{equation}
and
\begin{equation}\label{debye potential}
V_{\mbox{\tiny HTL}}(r)\propto\frac{e^{-q_i r}}{r},
\hspace{1cm}\mbox{with} \hspace{1cm} q_i=\sqrt{\frac{4\pi\alpha
T^2}{3}}.
\end{equation}

While on the completely one-loop dispersion curve, no pure
imaginary solution is found at $\omega\rightarrow 0$, which
implies the pole contains both real and imaginary parts and the
static potential takes the general form of damping oscillation
shown as Eq.(\ref{osc potential}). One can find out the poles
numerically by solving the equation
\begin{equation}\label{Famatt}
q^2-\frac{8\alpha}{\pi}\int^\infty_0 dp\
\frac{p^2}{E_p}\left[\frac{4E_p^2-q^2}{4p\ q}\log
\left(\frac{q-2p}{q+2p}\right)-1\right]n_f(E_p)=0,
\end{equation}
which is Eq.(\ref{define dispersion}) in the static limit
($\omega\rightarrow 0$) where the mode $q=q_r+iq_i$.

\begin{figure}
\resizebox{10cm}{!}{\includegraphics{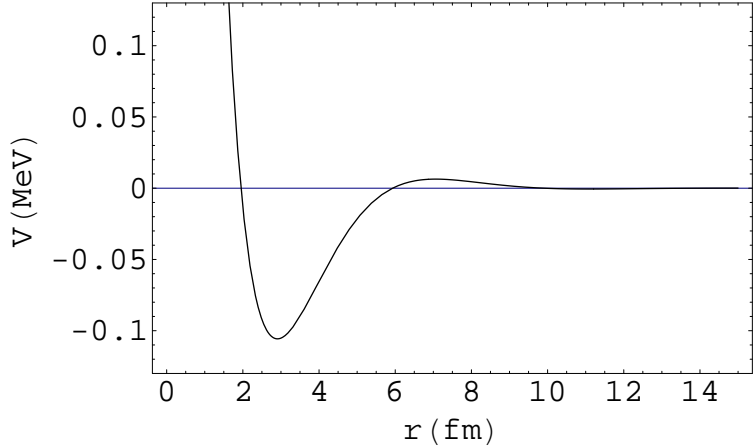}}
\caption{Oscillatory potential in completely one-loop calculation.
\label{fig osc potential}}
\end{figure}

In FIG.\ref{fig osc potential} we demonstrated the oscillatory
potential of QED. This damping oscillation is qualitatively
different for the monotonic Debye potential in Eq.(\ref{debye
potential}).

\section{radial distribution function and liquid QGP }

Generally speaking, In the picture of Debye screening, the
in-medium particles are "dressed" with the effective radii of
Debye length. Therefore the interactions among the component
particles are rather weak so that the system can be treated as the
ideal gas. However, once the Debye potential is replaced by the
oscillatory potential, the ideal gas is no longer a qualified
model. Then what kind of state of matter is the oscillatory
potential relevant to? To answer this question, one must know
about the typical character of each state.

\begin{figure}
%\begin{minipage}[t]{0.5\linewidth}
 %\centering
 \begin{center}
   \resizebox{10cm}{!}{\includegraphics{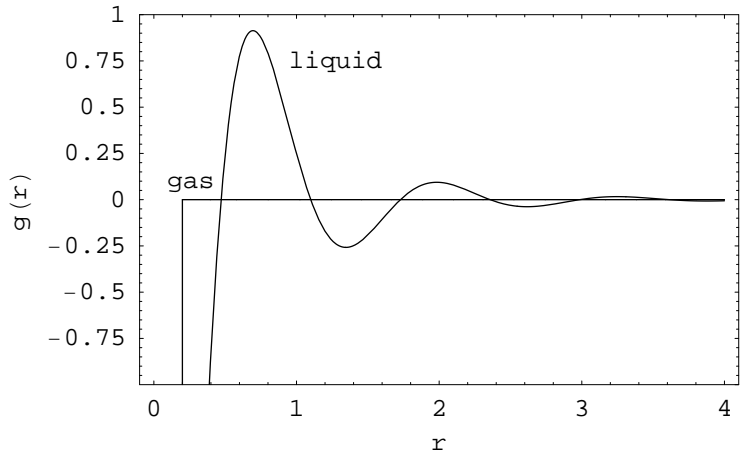}}
 \end{center}
\caption{\label{gas_liquid_RDF}Typical radial distribution
functions of gas and liquid.}
%\end{minipage}%
\end{figure}

To identify different states of matter, one is to distinguish the
different spacial configurations of the component particles. The
so-called radial distribution function (RDF), which is the
probability of finding two particles at a distance $r$ from each
other, is introduced as a powerful tool. For instance, particles
in the gas state are completely random, so that the possibilities
of finding any two particles are almost the same. Therefore its
RDF remains constant\footnote{The monotonic increasing is due to
the inaccessible core of the component particle.} as shown in
FIG.\ref{gas_liquid_RDF}. While the particles in the liquid state
have short range order so that the possibilities of finding nearby
particles are much larger than those far particles. Accordingly,
the RDF in the liquid state will present several damping peaks
along the radial direction\cite{March,Egelstaff,Ichimaru} which is
also sketched in FIG.\ref{gas_liquid_RDF}. We consider this
damping oscillation shape as the basic characteristic of a liquid
state, in other words, if someone could obtain such kind of RDF,
he may discover the footprint of a liquid state.
Thoma\cite{Thoma3} calculated the RDF of QGP in the HTL scheme,
which gives the exact Debye screening, and confirmed the negative
result for identifying a liquid. In the following, we will give up
the HTL scheme and work with complete one loop.

In the liquid state theory, one can define the RDF through
\begin{equation}\label{g-V}
g(r)=\exp\left[-{{V(r)}}/{ T}\right]
\end{equation}
where $V(r)$ is nothing else but the in-medium potential of
average inter-particle forces\cite{Egelstaff}. In the classical
liquid state theory, the RDF can be obtained analytically through
a certain pair potential model including the often used
Hypernetted-chain(HNC) or Percus-Yevick(PY) approximations, or
through some computer simulations like Monte Carlo or Molecular
dynamics\cite{March}. In this paper, we follow none of those
schemes, instead, we adopt the static in-medium potential obtained
in the completely one-loop calculation referring to the last
section.

\begin{figure}
%\begin{minipage}[t]{0.5\linewidth}
 %\centering
 \begin{center}
   \resizebox{10cm}{!}{\includegraphics{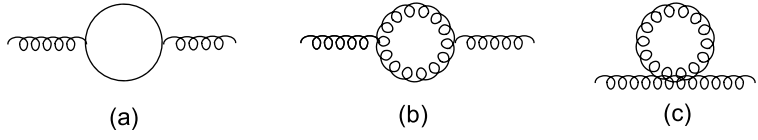}}
 \end{center}
\caption{\label{fig_qcd_pola} Gluon polarization.}
%\end{minipage}%
\end{figure}

As for plain QCD, the one-loop gluon polarization is determined by
the diagrams in FIG.\ref{fig_qcd_pola}. Compared with QED, QCD
involves the gluon self-coupling. One can calculate the
temperature-dependent polarization tensor in the framework of
thermal field theory, like what we do in the last section. We skip
the standard steps and directly present the expressions of
completely one-loop polarization tensor of QCD in the temporary
axis gauge (TAG) as\cite{Kapusta2}
\begin{eqnarray}
&&\hspace{-0.7cm}\Pi_L^{(a)}=\frac{8\alpha_s}{\pi}\int_0^\infty
dp\ \frac{p^2}{\omega_q}\left[\frac{4\omega_q^2-q^2}{4p\
q}\log\left(\frac{q-2p}{q+2p}\right)-1\right]n_f(\omega_q)\label{F_qcd_I}\\[0.4cm]
&&\hspace{-0.7cm}\Pi_L^{(b+c)}=-\frac{3\alpha_s}{\pi}\int_0^\infty
dp\
p\left\{4-\frac{2q^2}{p^2}+\frac{2p}{q}\left[1+\left(\frac{2p^2-q^2}{2p^2}\right)^2\right]
\log\left(\frac{q+2p}{q-2p}\right)\right\}n_b(p).\label{F_qcd_II}
\end{eqnarray}
$\omega_q=\sqrt{p^2+m_q^2}$ where $m_q$ is the quark mass.
$n_b(p)=(e^{\beta p}-1)^{-1}$ is the gluon distribution function.
Here we study the 2-flavor QGP. For the running coupling
$\alpha_s$, we use the two-loop renormalization group
expression\cite{Kaczmarek}
\begin{equation}
\alpha_s=\left[\frac{9}{2\pi}\ln\left(\frac{T}{\Lambda}\right)+\frac{16}{9\pi}
\ln\left(2\ln\left(\frac{T}{\Lambda}\right)\right)\right]^{-1},
\end{equation}
where $\Lambda=73$MeV for the temperature range $1\sim 2T_c$.

We would like to point out that although applying the linear
response theory to non-Abelian gauge theory is at the risk of
gauge noninvariance, the TAG is believed safe enough because in
this gauge one can obtain the same vacuum polarization corrected
effective charge as the renormalization group
charge\cite{Kapusta2}. We hope the discussion in TAG may give at
least the qualitative features of the potential and RDF.

\begin{figure}
%\begin{minipage}[t]{0.5\linewidth}
 %\centering
 \begin{center}
   \resizebox{10cm}{!}{\includegraphics{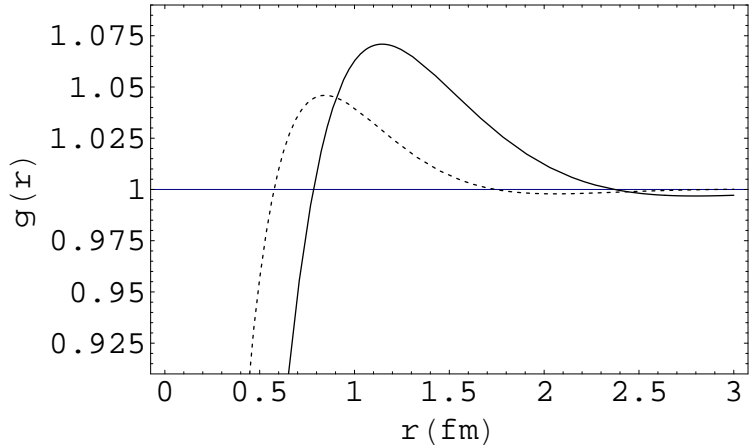}}
 \end{center}
%\end{minipage}%
\caption{\label{fig qcd RDF}RDF of QCD plasma. The solid and the
dotted lines are for $T$=0.2, 0.3GeV respectively.}
\end{figure}

Adding up Eqs.(\ref{F_qcd_I}) and (\ref{F_qcd_II}) and inserting
them into Eq.(\ref{potential}), one can find out the pole
numerically. Then the interquark potential (\ref{osc potential})
is obtained and so as to the RDF considering Eq.(\ref{g-V}).
FIG.\ref{fig qcd RDF} is the RDF of QCD plasma where we choose two
different temperatures 0.2 and 0.3GeV.\footnote{The deconfined QGP
is a Coulomb-like plasma, whose dimensionless coupling parameter
is $\Gamma=C_\alpha \alpha_s \left(\frac{3}{4\pi
n}\right)^{\frac{1}{3}}/T$ where $C_\alpha=4/3$ is the eignvalue
of the Casimir operator for quark and antiquark, $n$ is the
particle number density. For estimation, we take $n=6.3T^3$ for
2-flavor QGP by considering it as a massless
gas\cite{Gelman,Thoma4} For T=0.2 and 0.3GeV, the running coupling
constants are 0.5 and 0.35, and the corresponding coupling
parameters are 2.0 and 1.4, which are great than 1, indicating a
liquid state.} In FIG.\ref{fig qcd RDF}, one can see clearly the
damping oscillatory behavior of the RDF, which is very similar to
the typical shape of liquid in FIG.\ref{gas_liquid_RDF}. This
result might indicate the liquid state of hot QGP. Furthermore,
the RDF oscillation becomes weaker and weaker with the increase of
temperature, thus one may expect the QGP is approaching to an
ideal gas at the high temperature limit.

\section{discussion}

In this paper, we start with the comparison of dispersion
relations of HTL and complete one loop, pointing out an important
discrepancy in the dynamical screening regime which results in the
different behaviors of the static in-medium potentials. Then we
discuss the RDF of hot QGP. It appears an obvious damping
oscillation which implies the QGP might be in a liquid state.

How to deal with the interacting many-body system, especially the
strongly coupled or strongly correlated system, is a rather
difficult but fundamental problem. In principle, one can reduce
the many-particle distribution function to two- or single-particle
distribution function\cite{Mohling}. The RDF is actually the
two-particle distribution function. It is the basic physical
quantity in the atomic liquid theory that has been related to
various kinetic and thermodynamic
observables\cite{March,Egelstaff}. On one hand, the RDF is
obtained by considering certain dynamical and thermal statistical
model from the theoretical aspects. On the other hand, it can be
measured through scattering experiments in the atomic liquid.
Compare the theoretical RDF and the RDF extracted from
experiments, then one can figure out deeper discipline that rules
over the phenomenon. Parallel to the classical liquid theory, the
RDF in this paper is the static two-quark distribution function
with spherical symmetry. Although we can not measure the quark
distribution in QGP through scattering experiment as we do to the
atomic liquid, we can still measure the density-density
correlations, which is relevant to the Fourier transformation of
RDF\cite{Thoma3}, by observing the final state distributions. We
hope in this way, the picture in our calculation can be tested by
the experiments.

\begin{acknowledgments}
This work is partly supported by the National Natural Science
Foundation of China under project Nos. 10747135, 10675052 and
10575043.
\end{acknowledgments}

\end{document}